# Elicitation of SME Requirements for Cybersecurity Solutions by Studying Adherence to Recommendations

Alireza Shojaifar[1,2], Samuel A. Fricker[1,3], Martin Gwerder[1]

[1]FHNW, IIT and IMVS, 5210 Windisch, Switzerland
`(alireza.shojaifar|samuel.fricker|martin.gwerder)@fhnw.ch`
[2]Utrecht University, Dept. of Information and Computing Sciences, Utrecht, Netherlands
`a.shojaifar@uu.nl`
[3]Blekinge Institute of Technology, SERL-Sweden, 371 79 Karlskrona, Sweden
`samuel.fricker@bth.se`

**Abstract.** [Context and motivation] Small and medium-sized enterprises (SME) have become the weak spot of our economy for cyber attacks. These companies are large in number and often do not have the controls in place to prevent successful attacks, respectively are not prepared to systematically manage their cybersecurity capabilities. [Question/problem] One of the reasons for why many SME do not adopt cybersecurity is that developers of cybersecurity solutions understand little the SME context and the requirements for successful use of these solutions. [Principal ideas/results] We elicit requirements by studying how cybersecurity experts provide advice to SME. The experts' recommendations offer insights into what important capabilities of the solution are and how these capabilities ought to be used for mitigating cybersecurity threats. The adoption of a recommendation hints at a correct match of the solution, hence successful consideration of requirements. Abandoned recommendations point to a misalignment that can be used as a source to inquire missed requirements. Re-occurrence of adoption or abandonment decisions corroborate the presence of requirements. [Contributions] This poster describes the challenges of SME regarding cybersecurity and introduces our proposed approach to elicit requirements for cybersecurity solutions. The poster describes CYSEC, our tool used to capture cybersecurity advice and help to scale cybersecurity requirements elicitation to a large number of participating SME. We conclude by outlining the planned research to develop and validate CYSEC[1].

---

# 1 Introduction

Small and medium-sized enterprises (SME) are considered as an important part of economy specially e-driven economies [1, 2]. Browne et al. explain that based on EU commission 2005, any company with fewer than 250 employees and with annual turnover less than € 50 million can be considered as an SME [2]. Osborn with respect to EU Commission report states that these SME form 99% of European businesses [3]. Regarding cybercriminal point of view, the rate of cyber-attacks against SME is considerable [2, 4]. However, many SME (regarding Kaspersky Labs reports) do not believe and aware that they are the target of these attacks [1].

Browne et al. explain that SME are weaker targets than big companies since small companies have their own specific culture and behave differently regarding cybersecurity measures [2]. Kuusisto and Ilvonen explain that most of considered SME do not have documented information security policy, clearly determined security responsibilities and security training [5]. Kurpjuhn states that SME give primacy to business growth investment rather than security measures, however, the importance and severity of malicious threats in SME are the same as big companies, although the level of financial investment and resources for cybersecurity measures in SME are very low [6].

It should be noted that lack of investment and budget restrictions can be two main reasons of SME cybersecurity problems which can be originated from lack of security awareness by SME owners and lack of cost-effective processes [7, 8]. SME may also do not have an internal cyber security policy to reduce the possibility of cyber-attacks [7]. Xian et al. state that SME because of lack of budget, expertise and complexity of ISRAs (Information-security risk assessments) are not able to do ISRAs [9]. Gundu and Flowerday assert that some SME incline to concern about external threats and neglect the security risk of uninformed employees [10]. Also, small companies which may have low levels of risk tolerance can have different approach regarding perceived threats [2].

These are some of studied characteristics of SME, in turn, we can conclude that the cybersecurity approach which intend to safeguard SME against cyber-attacks should be different with large organisations. Different research vendors have considered and proposed some approaches, models or framework which address some of SME's characteristics [10, 8, 2]. Furthermore, Cholez and Girard concentrate on a method for SME's maturity assessment and process improvement in the context of information security management [11]. Mijnhardt et al. propose an assessment tool based on ISFAM (Information Security Focus Area Maturity) for information security advice for SME [12].

However, the preceding framework and models appear to consider some of SME characteristics, some of influential factors, or match particular SME (in specific country). ISFAM although covers different security areas in detail, seems complicated to apply regarding SME's level of knowledge and expertise in cybersecurity measures. More generally, many SME do not adopt good cybersecurity practices or abandon such practices for a variety of reasons, such as lack of information security knowledge and skill, lack of budget and resources, lack of security and risk awareness, and employees with multiple roles and access [8, 10]. Thus, although there are some attempts to alleviate the SME's cybersecurity problems, still a lack of understanding of the cybersecurity requirements of SME can be seen.

## 2  Requirements Elicitation by Studying Adherence

The here presented work aims at finding an effective way to elicit requirements of SME for cybersecurity solutions. The adoption of a cybersecurity recommendation hints at a correct match of the solution, hence successful consideration of the SME's context and requirements. Abandoned recommendations point to a misalignment that can be used as a source to inquire missed requirements. Such abandoning may be due to a variety of reasons that could point to requirements that are not satisfied by the cybersecurity solutions. Re-occurrence of adoption or abandonment decisions across many SME corroborate the presence of these requirements.

There are different requirements elicitation automated tools and feedback collection approaches such as Online ads and in-product surveys, Operational and event data, and A/B testing [16]. However, the point is that these approaches have not been applied and evaluated in the context of cybersecurity. The new idea of our approach is to study and mirror the approach of how cybersecurity experts provide advice to SME. The experts' recommendations offer insights into what important capabilities of the solution are and how these capabilities ought to be used for mitigating cybersecurity threats.

The study of adherence is performed by following the dialogue between a cybersecurity expert and the person in charge of cybersecurity in the SME. Such a dialogue may be structured according to the established plan-do-check-act (PDCA) model of process improvement [14] and be based on cybersecurity improvement frameworks like ISFAM [12]. We envision an incremental approach to cybersecurity improvement that matches the SME context where the customer is the priority and resources scarce. Inspired by agile development [15], we let the SME adopt cybersecurity capabilities that the person in charge ranks in a backlog of themes according to the perceived importance. Upon agreed timing, we let the cybersecurity expert and the SME review the achievements and reflect on successes and failures of adopting the cybersecurity controls. Table 1 outlines a cycle of this incremental improvement process. Although this cycle regarding cybersecurity problem can be the same for big organizations and SME, it can automate requirements elicitation for many more SME and we can have many more sources for requirements.

**Table 1.** Cycle of requirements elicitation by studying adherence to recommendations.

| Step | Elaboration | Example |
|---|---|---|
| 1 Recommend theme | The cybersecurity expert offers a portfolio of topics that could be relevant for the SME. | Offering of ISFAM focus areas, covering organisational, technical, and support themes for cybersecurity. |

| Step | Elaboration | Example |
|---|---|---|
| 2 Select cybersecurity theme | The person in charge of cybersecurity in the SME selects one of the cybersecurity topics suggested for improvement. | Choice of "secure software development." |
| 3 Recommend controls | The cybersecurity expert suggests a set of controls that offer protection against cyber attacks. | ISFAM-based suggestion of version control, source code and web scanning tools, defect management systems, and regression testing. |
| 4 Select controls to be adopted | The person in charge chooses among the controls and defines a date for review of the control implementation. | Choice of Git-based configuration and release management, Jenkins, and SonarQube with recommended security test cases. |
| 5 Monitor adherence | The cybersecurity expert and person in charge check adherence to the selected controls. | The cybersecurity expert and the person in charge identify inconsistent use of Git in the development activities, indicating that some employees did not use Git as intended. |
| 6 Obtain feedback | The cybersecurity expert elicits answers to questions like "what value did the selected control deliver?" or "why did you not use the control?" | Involvement of the employees that were not using Git as intended showed that they did not understand the release management process and perceived the tool to be too complicated. These feedback led to the replacement of git with GitKraken, a graphical client for Git that was well accepted even by junior developers. |

We expect that the study of feedback about adherence will be rich of insights that can be used to understand the requirements for cybersecurity solutions for SME. In the trials underlying Table 1, the SME identified controls, such as computer forensics, it was interested in and was not offered by the cybersecurity expert. Another feedback was that cybersecurity controls were offered that assumed an organisational structure that did not match the SME structure. Also, users had complained that too much unsolicited bulk emails arrive in their inboxes to initiate a change in the mail filters. For cybersecurity solution developers, these insights will be useful for planning new features or abandoning features that turn out to be unattractive. Some insights make explicit the validity of assumptions about the context of cybersecurity solution use, whether these assumptions were formulated explicitly or existed implicitly in the minds of the developers. Other insights offer concrete recommendations for how to adjust a control to make it useful in the SME context.

To reduce the cost of employing the method and allow scaling to many SME, we automate the dialogue and advice provision with a software that we call the Cybersecurity Coach (CYSEC). CYSEC allows cybersecurity experts to define themes and controls that they believe are helpful for SME. An SME can download and use CYSEC to determine its cybersecurity capability profile, obtain recommendations for improvement, and track the improvement success. Upon SME-defined timings, CYSEC encourages the SME to offer feedback about the selection decisions and the experience of implementing the selected practices. Consolidation of these observations and feedbacks across many SME will offer the community of cybersecurity developers and experts rich insights for evolving the solutions they are offering and advice they are suggesting.

CYSEC tool, in general, encompasses four different components: capability advisor, good practices and tools, adherence monitor, and a bot. Capability advisor regarding improvement model includes a questionnaire covering different cybersecurity capabilities (such as patch management, access control, …) referencing to good practices and SME can see their progress. Good practices and tools, provides SME with relevant information for training and tools for download. The adherence monitor as a goal monitor can help cybersecurity experts to evaluate their approaches. And the bot is an interactive element for Q&A through which each SME can receive feedback and suitable answers to their questions, and to realize the SME adherence to the advice. Through observing the SME adherence to the advice and evaluation by the cybersecurity experts, cybersecurity requirements elicitation for the SME can be done. However, although CYSEC has not developed yet and we aim to present the mock-up in the poster, the first three components are based on Duolingo's components (a successful tool in language learning). Moreover, we include a new component (advisory dialogue) for survey in the automated advice of SME to do requirements elicitation in the cybersecurity context.

We are developing the adherence monitoring-based requirements elicitation method with two organisations that are experts in cybersecurity and four SME that are interested in improving their cybersecurity capabilities. These four SME have guided us through development process with useful information and through the discussions we received interesting feedback regarding current frameworks. In about one year, we open the collaboration to further cybersecurity experts and SME with an open call for joining our work. The co-development approach allows us to understand the dialogue between the cybersecurity expert and the SME to the extent that it can be implemented in the CYSEC tool. The work with the experts allows us to understand important solutions and controls that are available and how they are used to address cybersecurity threats. The work with the SME offers us an opportunity to validate the requirements elicitation approach and mature the CYSEC tool.

## 3 Planned Research

Wieringa's Design Science framework [13] will be applied to conduct the research. The framework consists of a series of studies and actions that guide the design and validation of a method and tool like the CYSEC-enabled requirements elicitation approach. We emphasise the problem investigation, design and validation of the approach, and evaluation of the impact of the approach. To guide this work, our research aims to answer the following research questions:

RQ1. *What are the hurdles and enablers of SME to adopt cybersecurity solutions?*

RQ2. *Can the study of adherence to cybersecurity practice be used as a method of requirements elicitation for improving cybersecurity solutions?*

RQ3. *Can requirements elicitation be automated by embedding the dialogue between the cybersecurity expert and the person in charge of the SME in the CYSEC tool?*

RQ4. *What are the effects of the CYSEC tool-supported approach on cybersecurity capabilities of SME and solutions that support these SME?*

RQ1 will be answered by collecting experiences of the collaborating SME of using existing cybersecurity capability improvement methods. RQ2 will be answered by observing dialogues between cybersecurity experts and persons in charge of the collaborating SME from the perspective of requirements that can be identified in the dialogues. The results of RQ2 will be used for designing and implementing the CYSEC tool. RQ3 will be answered by iteratively letting SME use the CYSEC tool and evaluating whether the tool is understood and beneficial for the SME and whether the insights gained with the SME's end-user feedback helps the improve the cybersecurity solutions that were recommended to be used by CYSEC. RQ4 will be answered by inviting a larger number of SME to a beta evaluation phase of the CYSEC tool.

The outcome of the research will be an improved understanding of the requirements of solutions that protect SME against cyber threats. We expect that CYSEC as a tool not only improve the SME's adherence, knowledge, and awareness but also help cybersecurity experts with requirements elicitation for solutions that help SME to become secure.

## Acknowledgments

This work was made possible with funding from the European Union's Horizon 2020 research and innovation programme under grant agreement No 740787 (SMESEC) and the Swiss State Secretariat for Education, Research and Innovation (SERI) under contract number 17.00067. The opinions expressed and arguments employed herein do not necessarily reflect the official views of these funding bodies.